\documentclass[aps,pra,longbibliography,notitlepage,reprint,twocolumn,amsmath,amssymb,superscriptaddress]{revtex4-1}

\usepackage{graphicx}% Include figure files
\usepackage{bm}% bold math
\usepackage{braket}
\usepackage{natbib}
\setcitestyle{numbers,square,sort&compress}

\usepackage{here}

\usepackage[english]{babel}

\graphicspath{{pict/}{}}

\newcommand\hrefBibPDF[3][]{}

\usepackage{hyperref}
\hypersetup{pdfstartview={FitH},pdfpagemode={UseNone},
	colorlinks,linkcolor=blue, citecolor=blue, urlcolor=blue,
	bookmarksopen=true, pdfnewwindow=true}

\usepackage[all]{hypcap}

\usepackage[nodocdata=true]{pdfprivacy}

\begin{document}

\title{In-line detection and reconstruction of multi-photon quantum states}

\author{Kai~Wang}
\email[Corresponding author. Email: ]{kai.wang@anu.edu.au}
\affiliation{Nonlinear Physics Centre, Research School of Physics and
	Engineering, The Australian National University, Canberra, ACT 2601, Australia}
    \affiliation{Institute for Physics, Universit\"{a}t Rostock, Albert-Einstein-Str.~23, 18059, Rostock, Germany}

    \author{Sergey~V.~Suchkov}
\affiliation{Nonlinear Physics Centre, Research School of Physics and
	Engineering, The Australian National University, Canberra, ACT 2601, Australia}

\author{James~G.~Titchener}
\affiliation{Nonlinear Physics Centre, Research School of Physics and
	Engineering, The Australian National University, Canberra, ACT 2601, Australia}
\affiliation{Quantum Technology Enterprise Centre, Quantum Engineering Technology Labs, H. H. Wills Physics Laboratory and Department of Electrical and Electronic Engineering, University of Bristol, BS8 1FD, UK}

\author{Alexander~Szameit}
\affiliation{Institute for Physics, Universit\"{a}t Rostock, Albert-Einstein-Str.~23, 18059, Rostock, Germany}

\author{Andrey~A.~Sukhorukov}
\affiliation{Nonlinear Physics Centre, Research School of Physics and
	Engineering, The Australian National University, Canberra, ACT 2601, Australia}

\date{\today}

\begin{abstract}
Integrated single-photon detectors open new possibilities for monitoring inside quantum photonic circuits.
We present a concept for the in-line measurement of spatially-encoded multi-photon quantum states, while keeping the transmitted ones undisturbed.
We theoretically establish that by recording photon correlations from optimally positioned detectors on top of coupled waveguides with detuned propagation constants, one can perform
robust reconstruction of the density matrix describing the amplitude, phase, coherence and quantum entanglement.
We report proof-of-principle experiments using classical light, which emulates single-photon regime. Our method opens a pathway towards practical and fast in-line quantum measurements for diverse
applications in quantum photonics.
\end{abstract}

\maketitle

Quantum properties of multiple entangled photons underpin a broad range of applications~\cite{Zeilinger:2017-72501:PS} encompassing enhanced sensing, imaging, secure communications, and information processing.
Accordingly, approaches for measurements of multi-photon states are actively developing, from conventional quantum tomography with multiple bulk optical elements~\cite{James:2001-52312:PRA} to integrated photonic circuits~\cite{Shadbolt:2012-45:NPHOT}.
Latest advances in nanofabrication enable the integration of multiple single photon detectors based on superconducting nanowires~\cite{Natarajan:2012-63001:SCST, Kahl:2017-557:OPT}.

Beyond the miniaturization, integrated detectors open new possibilities for photon monitoring inside photonic circuits~\cite{Pernice:2012-1325:NCOM}.
However there remains an open question of how to perform in-line measurements of the quantum features of multi-photon states encoded in their density matrices, while ideally keeping the transmitted states undisturbed apart from weak overall loss. Such capability would be highly desirable similar to the classical analogues~\cite{Mueller:2016-42:OPT}, yet it
presents a challenging problem
since traditional approaches for quantum measurements are not suitable. In particular,
direct measurement methods~\cite{Lundeen:2012-70402:PRL, Thekkadath:2016-120401:PRL} are difficult to employ
due to complex measurement operators. On the other hand, the conventional state tomography~\cite{James:2001-52312:PRA}, a most widely used reconstruction method, requires reconfigurable elements to apply modified projective measurements in different time windows.
Recently, the reconstruction was achieved in static optical circuits~\cite{Titchener:2016-4079:OL, Oren:2017-993:OPT, Titchener:2018-19:NPJQI} or metasurfaces~\cite{Wang:1804.03494:ARXIV}, however it comes at a cost of spreading out quantum states to a larger number of outputs, which is incompatible with in-line detection principle.

\begin{figure}[t]
\centering
\includegraphics[width=\linewidth]{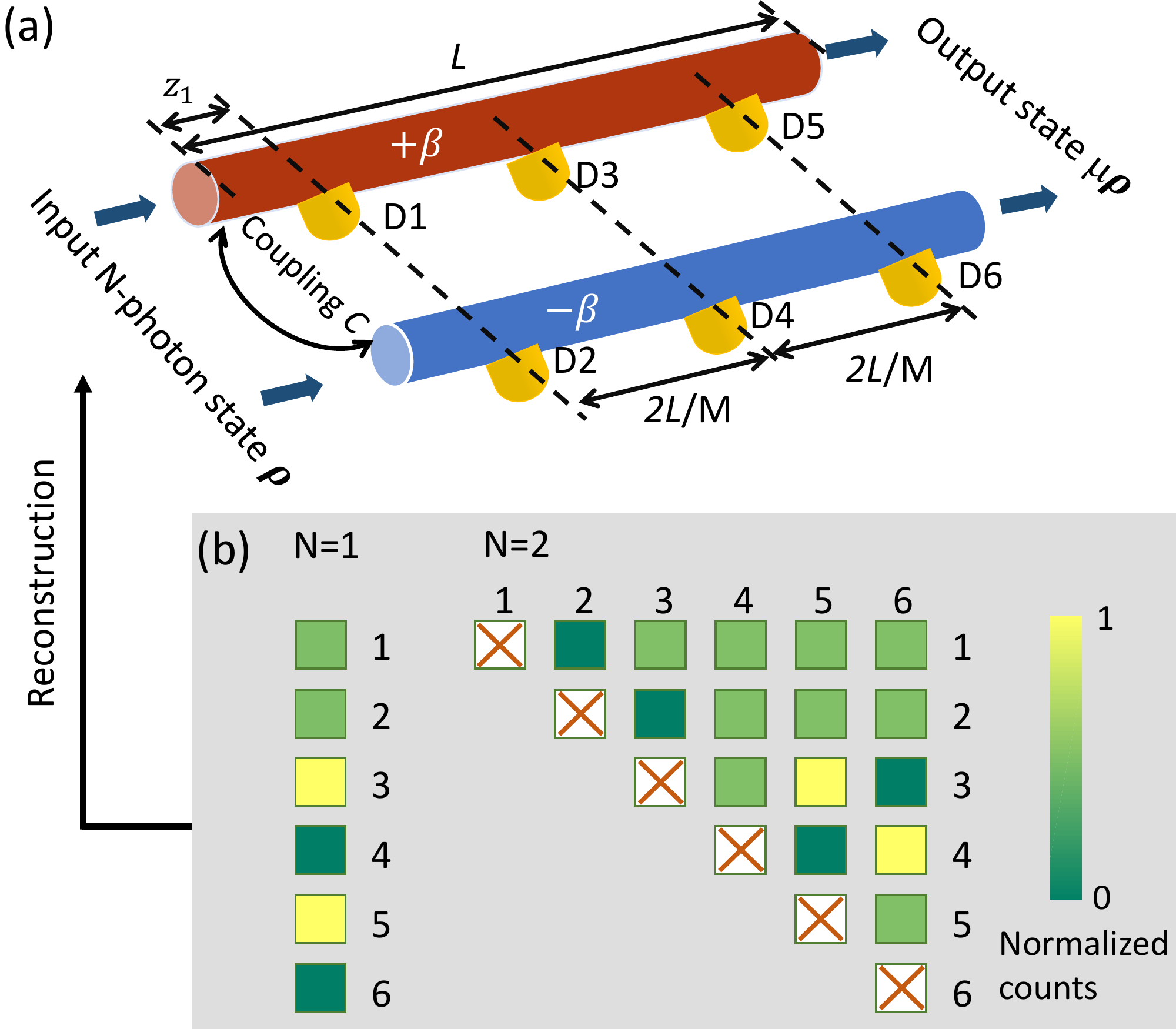}
\caption{Conceptual sketch of in-line detection and reconstruction of $N$-photon state ${\rho}$ with two spatial modes. (a)~Two waveguides with coupling constant $C$ and detuning $\beta$ in propagation constants.
An even number $M$ of single-photon click detectors are positioned at $M/2$ equidistant cross-sections, illustrated for $M=6$ with D1--D6 labels. (b)~Examples of simulated single-photon counts ($N=1$) and $N$-fold correlations (for $N=2$) for $z_1=0$ and $\beta=C / \sqrt{2}$, which enable full reconstruction of the input density matrix $\rho$.}
\label{fig1}
\end{figure}

In this work, we present a new conceptual approach for practical in-line measurement of multi-photon states using integrated detectors which is suitable
for multi-port systems. We illustrate it for two waveguides in Fig.~\ref{fig1}(a). The  coupled waveguides (coupling coefficient $C$) have different propagation constants (detuned by $\beta$) and the length $L$ is exactly a revival period. An even number $M$ of weakly coupled single-photon click detectors are placed at $M/2$ cross-sections, starting from $z_1$ with equal distances $2L/M$ between each other. As we demonstrate in the following, the measurement of $N$-fold nonlocal correlations by averaging the coincidence events enables a full reconstruction of the density matrix $\rho$ for $N$-photon states.

\begin{figure}[bt]
\centering
\includegraphics[width=\linewidth]{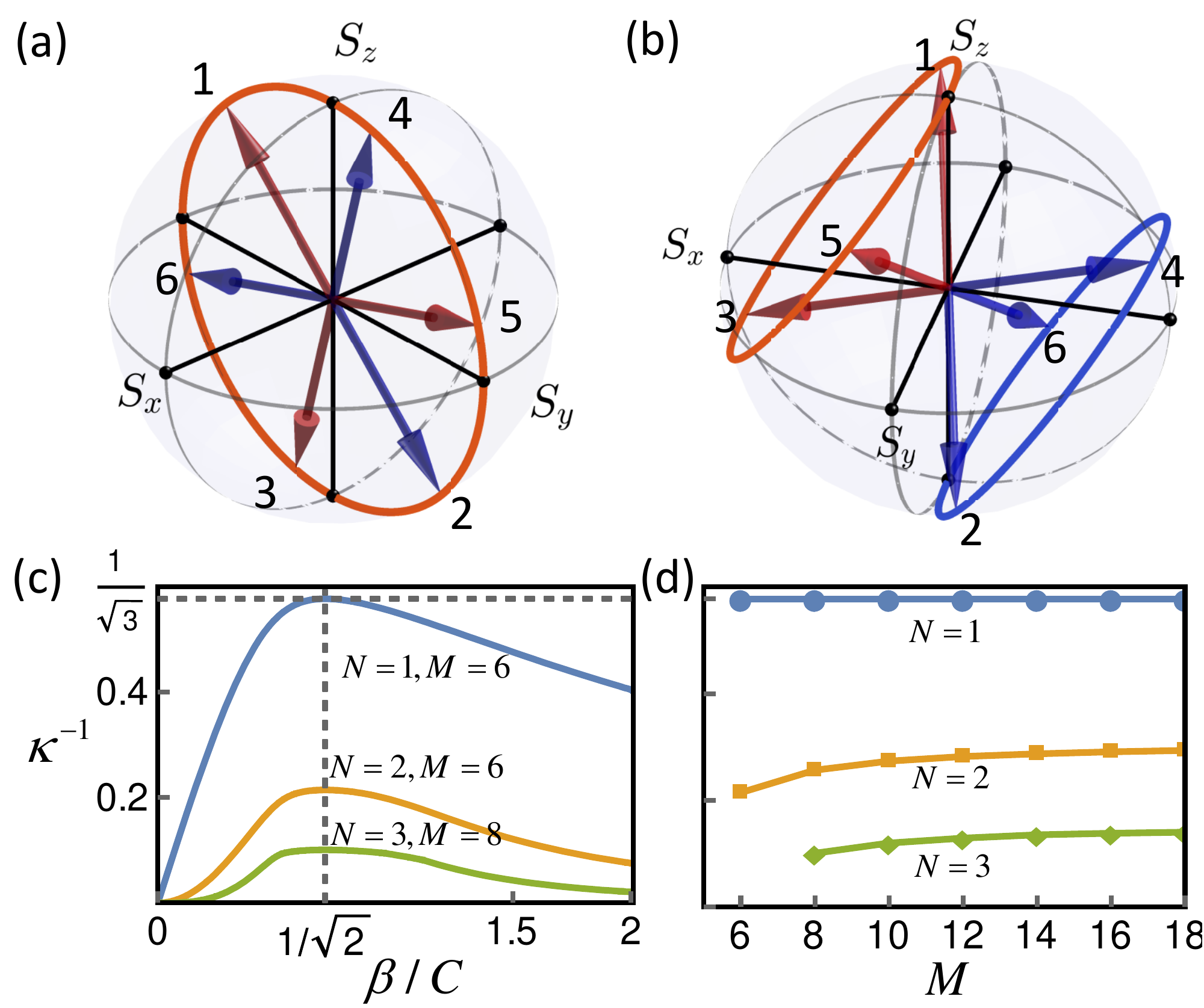}
\caption{Robust reconstruction with waveguide detuning. (a,b)~The analysis states on a Bloch sphere for (a)~identical waveguides ($\beta=0$) and (b)~introduced detuning ($\beta=C/\sqrt{2}$). Curves show the evolution along $z$, and arrows correspond to six detectors as indicated in Fig.~\ref{fig1}(a). Red and blue colors denote waveguide a and b, respectively. (c,d)~Inverse condition number $\kappa^{-1}$ versus (c)~normalized detuning $\beta/C$ and (d)~number of detectors $M$ for the optimal detuning $\beta=C/\sqrt{2}$.
}
\label{fig2}
\end{figure}

The quantum state evolution along the waveguides, under the assumption of weakly coupled detectors, is governed by the Hamiltonian
\begin{equation}
   \hat{H}= \beta\hat{a}_1^\dagger\hat{a}_1
            -\beta\hat{a}_2^\dagger\hat{a}_2
            +C\hat{a}_1^\dagger\hat{a}_2 + C\hat{a}_2^\dagger\hat{a}_1,
\end{equation}
where $\hat{a}_{q}^\dagger$ ($\hat{a}_{q}$) are the photon creation (annihilation) operators in waveguide number $q$, $\beta$ describes the propagation constant detuning, and $C$ is the coupling constant. We use Heisenberg representation~\cite{Bromberg:2009-253904:PRL, Peruzzo:2010-1500:SCI} and map time $t$ to the propagation distance $-z$. Hence the operator evolution reads
\begin{equation}
  \hat{a}_q(z) = \sum_{q'=1,2} \mathbf{T}^{\ast}_{q,q'}(z) \hat{a}_{q'}(0), \quad
  \hat{a}^\dagger_q(z) = \sum_{q'=1,2} \mathbf{T}_{q,q'}(z) \hat{a}_{q'}^\dagger(0).
\end{equation}
Here the linear transfer matrix elements of the waveguide coupler are
$\mathbf{T}_{q,q}(z) = \cos(\eta z) + i (-1)^{q} {\beta}{\eta^{-1}}\sin(\eta z)$ and $\mathbf{T}_{q,3-q} = -i{C}{\eta^{-1}}\sin(\eta z)$,
where $\eta = (C^2 + \beta^2)^{1/2}$. We note that the input state is restored at the revival length $L = 2 \pi / \eta$, since $\hat{a}^\dagger_q(L) = \hat{a}^\dagger_q(0)$. While the in-line detectors introduce small loss, due to their symmetric positions the revival effect will remain, and the output density matrix would only exhibit an overall loss $\mu \rho$, where $\mu$ can be close to unity.

We consider the measurement of states with a fixed photon number $N$, which is a practically important regime~\cite{James:2001-52312:PRA, Titchener:2016-4079:OL, Oren:2017-993:OPT, Titchener:2018-19:NPJQI}. We base the analysis on the most common type of click detectors that cannot resolve the number of photons arriving simultaneously at the same detector, and cannot distinguish which photon is which.
Then, the measurement of a single photon by detector number $m$ can be described by an operator $\hat{A}_m = \hat{a}_{q_m}^\dagger(z_m) \hat{a}_{q_m}(z_m)$, where $q_m$ is the waveguide number and $z_m$ is the coordinate along the waveguide at the detector position.
Accordingly, the photon correlations corresponding to the simultaneous detection of $N$ photons by a combination of $N$ detectors with numbers $m_1, m_2, \ldots, m_N$ are found as:
\begin{equation} \label{eq:correl}
  \Gamma(m_1, m_2, \ldots, m_N) = {\rm Tr}( \rho \hat{A}_{m_1} \hat{A}_{m_2} \cdots \hat{A}_{m_N} ) .
\end{equation}
For a single photon ($N=1$), the measurement corresponds to a direct accumulation of counts at each detector without correlations, and we show an example in Fig.~\ref{fig1}(b, left) for a pure input state $\ket{\psi_i}=[1,1]^\mathrm{T}/\sqrt{2}$ with the density matrix $\rho = \ket{\psi_i}\bra{\psi_i}$.
We present an example of coincidence counts for a two-photon N00N state ($N=2$) in Fig.~\ref{fig1}(b, right), corresponding to the events of clicks at different combinations of two detectors.

We outline the principle of input $N$-photon density matrix $\rho$ reconstruction from the correlation measurements.
The vectors $\ket{\psi_q}=[\mathbf{T}_{q,1}, \mathbf{T}_{q,2}]^\mathrm{T}$ are the analysis states, which define the measurements at the different detector positions.
We visualize their evolution along $z$ on a Bloch sphere, where the projector $\ket{\psi_q}\bra{\psi_q}$ is decomposed into the Pauli matrices $\hat{\sigma}_i\ (i=x,y,z)$ to generate the coordinates $S_i=\mathrm{Tr}(\hat{\sigma}_i \ket{\psi_q}\bra{\psi_q})$. First
we present the trajectories for the case of no detuning ($\beta=0$) in Fig.~\ref{fig2}(a),
where we see that the analysis states simply trace out a circle along the Bloch sphere. Considering for example $M=6$ detectors as sketched in Fig.~\ref{fig1}(a), all the analysis states corresponding to detectors shown by arrows
are in one plane. Such a configuration cannot probe states beyond that plane, therefore one cannot perform state reconstruction with a static, non-detuned directional coupler. In contrast, for detuned waveguides ($\beta \ne 0$) the circular trajectories in the first and second waveguides become non-degenerate, see the red and blue curves in Fig.~\ref{fig2}(b). We indicate the analysis states with arrows for $M=6$, and note that they are spread out to different directions in the sphere and can thus be utilized to probe all information about the states and enable the density matrix reconstruction.

The input state reconstruction is performed as follows.
First, we introduce an index $p = 1, 2, \ldots, P$ with $P=M!/[N!(M-N)!]$ to enumerate all possible $N$ combinations of $M$ detectors, $(m_1, m_2, \ldots, m_N)_p$. Second, we select $S=(N+3)!/(N! 3!)$ real-valued parameters $r_1, r_2, \ldots, r_S$, which represent independent real and imaginary parts of the density matrix elements following the procedure described in the Supplementary of Ref.~\cite{Titchener:2018-19:NPJQI}. Here, we consider the indistinguishable detection scheme that does not tell which photon is which. Then, we reformulate Eq.~(\ref{eq:correl}) in a matrix form:
\begin{equation} \label{eq:matrix}
  \Gamma_p \equiv \Gamma(m_1, m_2, \ldots, m_N)_p = \sum_{s=1}^S {\bf B}_{p,s} \, r_{s} ,
\end{equation}
where the elements of matrix ${\bf B}$ are expressed through the transfer matrices $\mathbf{T}$ at the detector positions.
The density matrix parameters $\{r_s\}$ can be reconstructed from the correlation measurements by performing pseudo-inversion of Eq.~(\ref{eq:correl}) provided $P \ge S$, i.e. when the number of detectors is
\begin{equation} \label{eq:Mmin}
    M \ge N+3 .
\end{equation}

We now analyze the robustness of reconstruction with respect to possible experimental inaccuracies in the correlation measurements, such as shot noise. This can be quantified by the condition number $\kappa$ of the matrix ${\bf B}$ \cite{Foreman:2015-263901:PRL, Titchener:2018-19:NPJQI}. Then, the most accurate reconstruction corresponds to the smallest condition number.
We numerically calculate the condition numbers for different combinations of $M$ and $N$, considering a symmetric arrangement of detectors as sketched in Fig.~\ref{fig1}(a). We
find that accurate reconstruction can be achieved for the number of click detectors $M > N+3$. However, the condition number is infinite and the reconstruction cannot be performed for $M = N+3$.
This happens because
the $M$ analysis states are not fully independent, but actually constitute $M/2$ pairs of orthogonal states, as illustrated in Fig.~\ref{fig2}(b).
There is no contradiction with Eq.~(\ref{eq:Mmin}), since it establishes only a necessary condition for reconstruction.

We determine that a detuning of the waveguide propagation constants ($\beta$) is essential to perform the reconstruction. We illustrate in Fig.~\ref{fig2}(c) that the variation of $\beta$ strongly changes the reconstruction condition for a different number of photons $N=1,2,3$, considering the minimum possible even number of detectors. With no detuning, for $\beta=0$, $\kappa^{-1}$ goes to zero, meaning that the reconstruction is ill-conditioned and cannot be performed in practice. The optimal measurement frames occur at $\beta/C \simeq 1/\sqrt{2}$. In Fig.~\ref{fig2}(d) we fix $\beta/C=1/\sqrt{2}$ and plot $\kappa^{-1}$ versus the number of detectors $M$.
We find that $\kappa = 1 / \sqrt{3}$ for $N=1$ and any $M \ge 6$, in agreement with classical measurement theory \cite{Foreman:2015-263901:PRL}. For multi-photon states, the inverse condition number slightly increases for a larger number of detectors.
This happens because we consider click detectors which do not resolve multiple photons. The reconstruction is possible even in absence of measurement counts corresponding to several photons arriving to the same detector, yet as for larger number of detectors such events become less probable, the reconstruction accuracy might be slightly improved. However this improvement could be offset by a higher number of dark counts, so the optimal detector number will depend on their characteristics.

\begin{figure}[t]
\centering
\includegraphics[width=\linewidth]{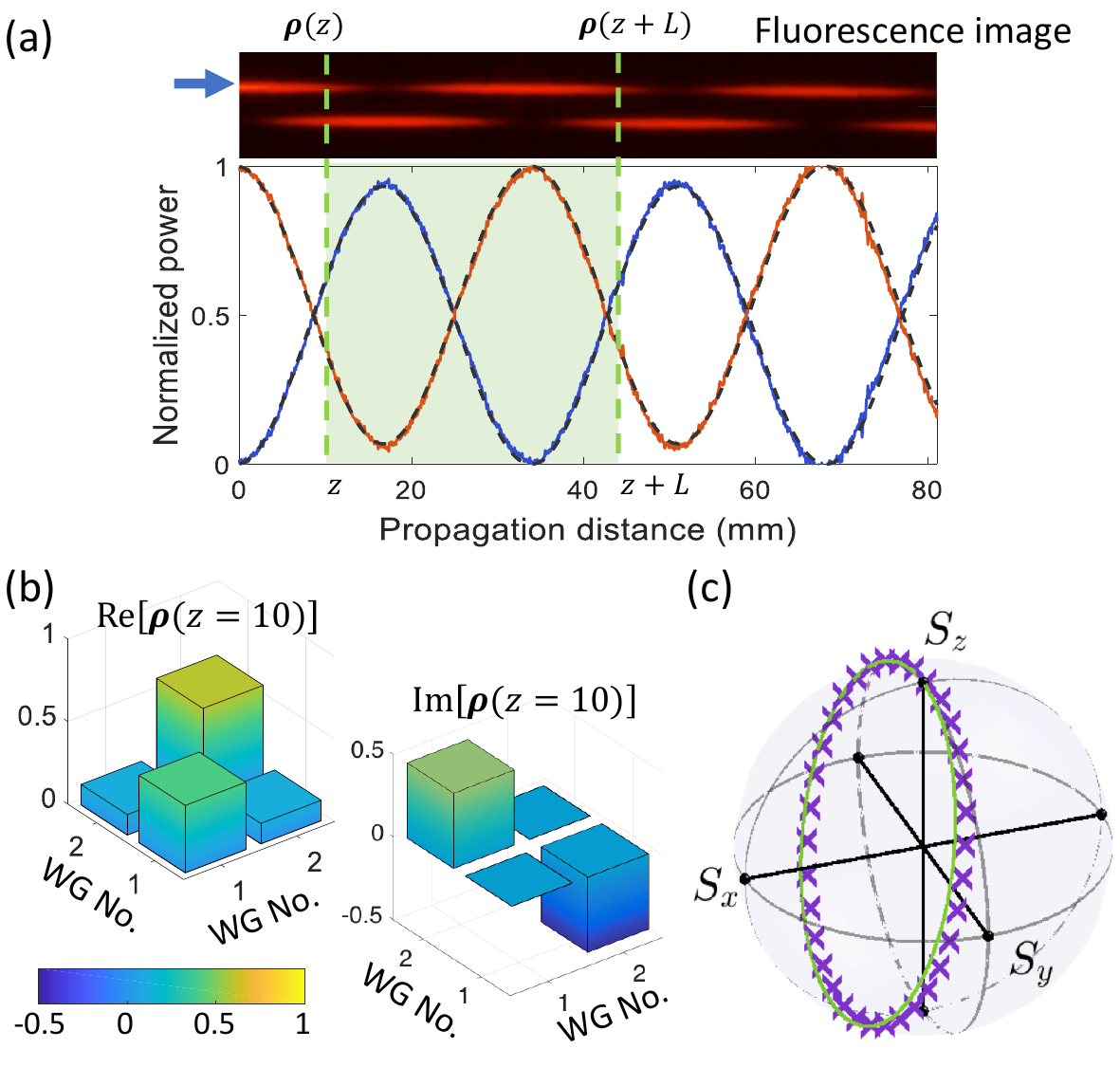}
\caption{Experimental in-line measurement and reconstruction of $N=1$ states emulated with classical laser light in detuned directional couplers.
(a)~A fluorescence image (top), showing light intensity along the coupler. Bottom~-- the corresponding normalized power evolution in each waveguide (solid curves) compared to theory (dashed curves).
Shading indicates a section $(z,z+L)$, where in-line measurements are used to reconstruct the full density matrix at the section input, $\rho(z)$.
(b)~The reconstructed real and imaginary parts of the density matrix elements for different combinations of the waveguide numbers (WG No.) at $z=10\ \mathrm{mm}$, the beginning of the shaded section in (a).
(c)~Experimentally reconstructed $\rho(z)$ from a set of different positions $z$ (crosses) compared with theoretical predictions (green curve) on a Bloch sphere, demonstrating an average fidelity of $99.65\%$.
}
\label{fig3}
\end{figure}

We demonstrate the scheme with {\em proof-of-principle experiments} for classical light. Mathematically, this is equivalent to the single-photon problem~\cite{Perets:2008-170506:PRL, Peruzzo:2010-1500:SCI}, since monochromatic laser light can be described by a density matrix of the same dimensionality as for a single photon, $2 \times 2$ corresponding to two waveguides.
We employ laser-written waveguides in fused silica to perform a reconstruction with a source of coherent laser light. In the fabrication process, we use different laser energies to write the coupled waveguides, and thereby achieve a pre-determined offset ($\pm\beta$) of the propagation constants for the fundamental modes in the two waveguides~\cite{Szameit:2010-163001:JPB}. We launch 633~nm laser light into one of the waveguides, and perform in-line measurements of  the evolution of intensity by observing the fluorescence (around 650 nm wavelength) emitted from the color centers of the glass material under a microscope~\cite{Szameit:2007-241113:APL, Szameit:2010-163001:JPB}. We present a characteristic fluorescence image in Fig.~\ref{fig3}(a, top), which features a clear beating pattern due to light coupling between the waveguides along the propagation direction $z$. The extracted normalized power in each waveguide is shown in Fig.~\ref{fig3}(a, bottom). We observe a good agreement with the theory (dashed line) for the coupled waveguide modes with $C=0.0885\ \mathrm{mm^{-1}}$ and propagation constants detuning $\beta=0.0240\ \mathrm{mm^{-1}}$.
This confirms that the fluorescence indeed acts as in-line detection, which is conceptually equivalent to putting many weakly coupled detectors homogeneously along both waveguides.

We then test our reconstruction method using the fluorescence image of the waveguide coupler. In the 80-mm long section of the fabricated coupler shown in Fig.\ref{fig3}(a), light periodically couples several times between the waveguides over this distance.
We truncate a section along $z$ with one revival period $L$ to mimic the in-line quantum detection. Then, the state $\rho(z)$ can be predicted by the coupled wave equation using the characterized parameters as described above. We consider different $z$ as starting positions, which allows us to effectively change the input state, and verify it's reconstruction accuracy from the fluorescence image in the truncated section $(z,z+L)$. We employ a maximum-likelihood method to perform a pseudo-inversion of Eq.~(\ref{eq:matrix}) and thereby find an input density matrix that best fits the evolution of the fluorescence power in the two waveguides, see an example in Fig.~\ref{fig3}(b). We analyze 34 different input states, truncated from successive $z$ with 1~mm increments. The reconstructed states are plotted on a Bloch sphere with crosses in Fig.~\ref{fig3}(c), and compared to theoretical modelling results shown with a green curve. The coordinates are obtained via decomposing the density matrices to the Pauli matrices, i.e. $S_i=\mathrm{Tr}(\hat{\sigma_i}\rho)$ with $i$ spanning $x,y,z$. We observe an excellent consistency between direct theoretical modelling and reconstruction from the experimental fluorescence images of the full density matrix, where we reach a very high average fidelity of $99.65\%$.

\begin{figure}[tbp]
\centering
\includegraphics[width=\linewidth]{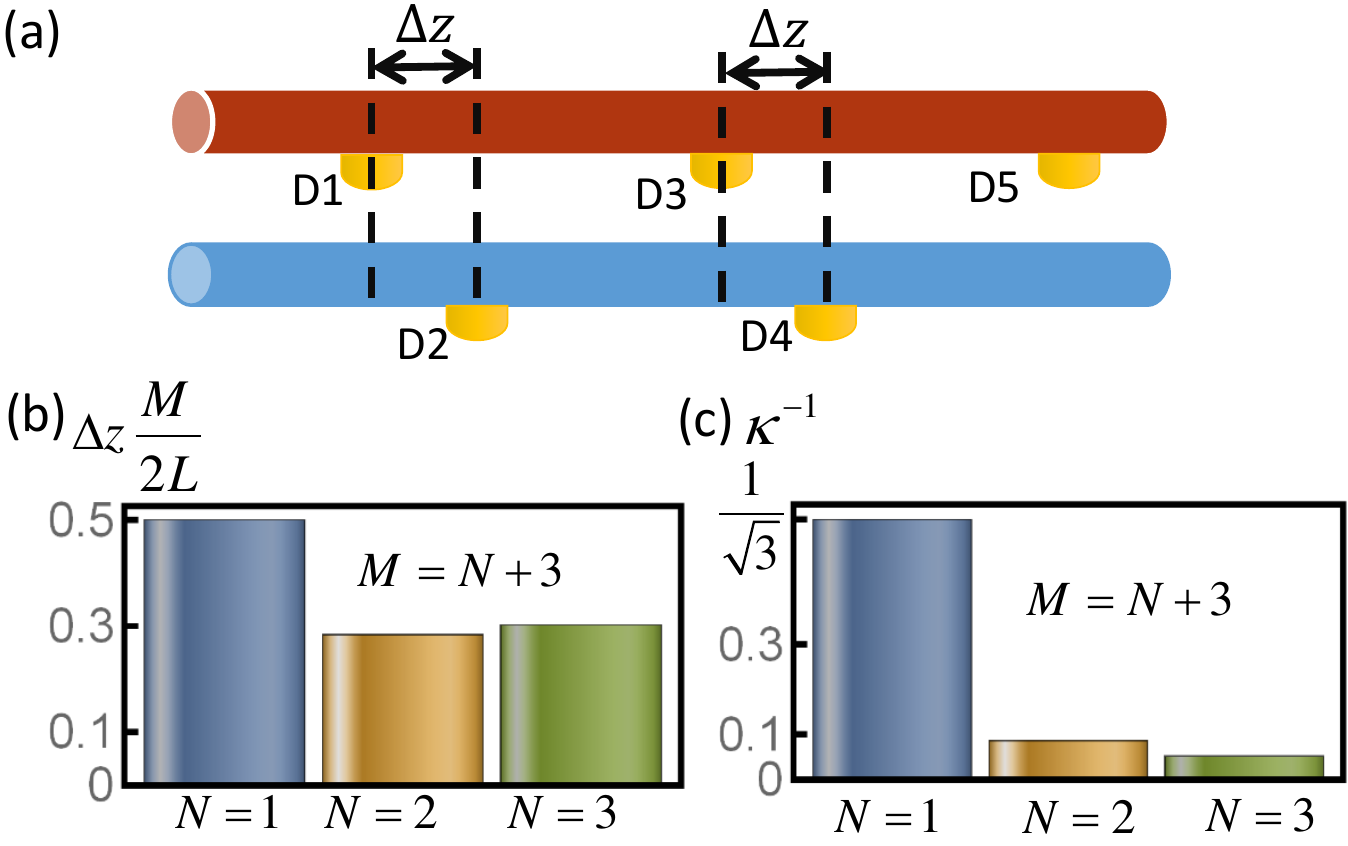}
\caption{Minimization of detector numbers to the limit $M=N+3$: (a) Conceptual sketch of introducing a shift $\Delta z$ for all detectors in one waveguide and skip the last detector if $M$ is an odd number. (b) Normalized $\Delta z$ (to $2L/M$) that achieves the best reconstruction condition with $\beta=C/\sqrt{2}$ for $N=1,2,3$. (c) The corresponding inversed condition number for the three cases in (b).}
\label{fig4}
\end{figure}

Finally, we explore the possibility to perform reconstruction with the minimal number of detectors, $M=N+3$ according to Eq.~(\ref{eq:Mmin}). We perform extensive numerical modelling by optimizing $\beta$ and the individual positions of all detectors. Although asymmetric positions may lead to a slight modification of the output state, this could be minimized by reducing the detector coupling.
We find that well-conditioned reconstruction with $M=N+3$ can be achieved for $N=1,2,3$ with the detung $\beta/C=1/\sqrt{2}$ and by shifting all detectors in one waveguide by $\Delta z$, as sketched in Fig.~\ref{fig4}(a) for $N=2$ and $M=5$.
We present in Fig.~\ref{fig4}(b) the optimal shifts $\Delta z$, normalized to the separation $2L/M$ between neighboring detectors in a waveguide. The corresponding inverse condition numbers are shown in Fig.~\ref{fig4}(c). For $N=1$ the best case appears at $\Delta z M/2L=0.5$, which means the four detectors are arranged in a zig-zag manner, giving rise to an optimal inverse condition number $\kappa^{-1}=1/\sqrt{3}$, the same as for a larger number of detectors as shown in Fig.~\ref{fig2}(d). For multi-photon states with $N=2,3$, the optimal values of $\Delta z$ correspond to asymmetric detector positions, while the inverse condition numbers are lower (worse) compared to larger detector numbers [c.f. Fig.~\ref{fig2}(d)].
We note that for higher photon numbers $N \ge 4$, reconstruction with $M=N+3$ appears impossible for any detector positions. We anticipate that this limitation can be overcome by modulating the waveguide coupling and detuning along the propagation direction, which would require further investigation beyond the scope of this work.

In conclusion, we proposed a practical and efficient approach for in-line detection and measurement of single and multi-photon quantum states in coupled waveguides with integrated photon detectors, suitable for various applications in quantum photonics. We showed proof-of-principle results with laser-written waveguides for classical light emulating single-photon regime.
We presented the theory and experiments for a two-port system, and this can be scaled to multiple coupled waveguides.
Moreover, our approach has a potential for translation from spatial to frequency and time-domain measurements.

We gratefully thank financial support from Australian Research Council (DP160100619); the Australia-Germany Joint Research Cooperation Scheme; Erasmus Mundus (NANOPHI 2013 5659/002-001); the Alexander von Humboldt-Stiftung; German Research Foundation (SZ 276/9-1, SZ 276/12-1, BL 574/13-1, SZ 276/15-1, SZ 276/20-1).

\end{document}